\documentclass[conference]{IEEEtran}
\IEEEoverridecommandlockouts
\usepackage{cite}
\usepackage{amsmath,amssymb,amsfonts}
\usepackage{algorithmic}
\usepackage{graphicx}
\usepackage{textcomp}
\usepackage{xcolor}
\usepackage{todonotes}
\def\BibTeX{{\rm B\kern-.05em{\sc i\kern-.025em b}\kern-.08em
    T\kern-.1667em\lower.7ex\hbox{E}\kern-.125emX}}
\begin{document}

\title{APIKS: A Modular ROS2 Framework for Rapid Prototyping and Validation of Automated Driving Systems\\

}


\author{\IEEEauthorblockN{Jo\~{a}o-Vitor Zacchi, Edoardo Clementi, Núria Mata}
\IEEEauthorblockA{
\textit{Fraunhofer Institute for Cognitive Systems (IKS)}\\
Munich, Germany \\
\{firstname.lastname\}@iks.fraunhofer.de}
}

\maketitle

\begin{abstract}

Automated driving technologies promise substantial improvements in transportation safety, efficiency, and accessibility. However, ensuring the reliability and safety of Autonomous Vehicles in complex, real-world environments remains a significant challenge, particularly during the early stages of software development. Existing software development environments and simulation platforms often either focus narrowly on specific functions or are too complex, hindering the rapid prototyping of small proofs of concept. To address this challenge, we have developed the APIKS automotive platform, a modular framework based on ROS2. APIKS is designed for the efficient testing and validation of autonomous vehicle software within software-defined vehicles. It offers a simplified, standards-based architecture designed specifically for small-scale proofs of concept. This enables rapid prototyping without the overhead associated with comprehensive platforms. We demonstrate the capabilities of APIKS through an exemplary use case involving a Construction Zone Assist system, illustrating its effectiveness in facilitating the development and testing of autonomous vehicle functionalities.


\end{abstract}

\section{Introduction}
\label{sec:intro}
The rapid advancement of autonomous vehicle (AV) technologies holds the promise of transforming transportation systems by enhancing safety, efficiency, and convenience. Autonomous vehicles have the potential to significantly reduce traffic accidents caused by human error \cite{wangQuantificationSafetyImprovements2024}, optimize traffic flow to alleviate congestion, and provide accessible transportation options for those unable to drive. They can revolutionize logistics and public transportation, leading to economic benefits and improved quality of life. However, ensuring the reliability and safety of AV software in complex and dynamic environments remains a significant challenge. These vehicles must navigate unpredictable real-world conditions, interpret vast amounts of sensory data in real-time, and make critical decisions that should prioritize the safety of all road users. Comprehensive testing and validation are essential to address these challenges, necessitating tools and frameworks that can efficiently simulate a wide array of real-world scenarios. Such simulations must encompass diverse environmental conditions, traffic patterns, and potential hazards to thoroughly evaluate the performance and safety of AV systems before deployment.

Simulation tools have become indispensable in the development and testing of AV systems. They enable extensive testing across diverse conditions much faster than real-world trials, achieving speeds up to $10^3$ to $10^5$ times faster \cite{fengDenseReinforcementLearning2023}. Existing simulators often emphasize specific functions—such as vehicle dynamics, driving policy development, traffic simulation, or sensory data generation. However, there is a scarcity of platforms offering an integrated approach for end-to-end AV testing that also supports rapid prototyping and streamlined development.

Comprehensive platforms like Apollo \cite{baiduApolloOpenAutonomous} and Autoware \cite{autowarefoundationAutoware} provide robust solutions capable of deployment in actual vehicles. While these platforms are invaluable for full-scale AV development, their complexity can hinder rapid testing and validation of small proofs of concept. The extensive feature sets and deployment capabilities introduce overhead that may not be necessary for early-stage development, making it challenging to quickly iterate and test new ideas.

In response to these challenges, we introduce the APIKS\footnote{Following the double-blind process, we cannot divulge the meaning of the accronym} automotive platform, a modular framework based on Robot Operating System 2 (ROS2) middleware \cite{macenskiRobotOperatingSystem2022} designed for the efficient testing and validation of autonomous vehicle software. APIKS offers a simpler architecture tailored for small proofs of concept, enabling rapid prototyping without the complexities associated with full-scale deployment platforms. By basing our development on industry standards, including alignment with the ISO/TR 4804 guidelines \cite{isoRoadVehiclesSafety2020}, APIKS builds on automotive best practices in safety and reliability. The framework streamlines development by being modular and software-centric, allowing developers to focus on specific components and functionalities as needed.

Our contributions are the following:

\begin{itemize}
    \item We present APIKS, a modular framework that facilitates efficient testing and validation of AV software, particularly suited for rapid prototyping and small proofs of concept.
    \item We showcase the potential of APIKS in a Construction Zone Assist (CZA) use case, illustrating its effectiveness in enabling the deployment and testing of new automated driving functionalities in a simulated environment.
\end{itemize}

The remainder of this paper is organized as follows: Section~\ref{sec:background} provides a background on the core concepts behind APIKS. Section~\ref{sec:framework} outlines the details of our proposed platform. Section~\ref{sec:cza} 
introduces the Construction Zone Assist, an exemplary use case of an automated vehicle function deployed in APIKS. Finally, Section~\ref{sec:conclusion} concludes the paper by summarizing our contributions and discussing potential avenues for future work in this domain.

\section{Background}
\label{sec:background}

\subsection{ROS2 and ADS Midlewares}
Middleware solutions are crucial for Automated Driving Systems (ADS) software, providing the communication infrastructure necessary for seamless data exchange between various components. ROS2 has emerged as a leading middleware framework by leveraging the Data Distribution Service (DDS) \cite{pardo-castelloteOMGDataDistributionService2003} for decentralized communication, offering enhanced scalability, real-time performance, and security compared to ROS 1 \cite{quigleyROSOpensourceRobot}.

In ROS2, topics, publishers, and subscribers form the core of its publish-subscribe communication model:
\begin{itemize}
    \item \textbf{Topics} are named channels that facilitate the transmission of data between nodes.
    \item \textbf{Publishers} are nodes that send messages to specific topics.
    \item \textbf{Subscribers} are nodes that receive messages from topics they are interested in.
\end{itemize}

This structure allows multiple nodes to asynchronously exchange information efficiently. Additionally, ROS2 supports \emph{services} for synchronous, request-response interactions between nodes. Services enable one node to request specific actions or information from another, ensuring direct and immediate communication when needed.

Autoware \cite{autowarefoundationAutoware} operates on the ROS 2 middleware framework and serves as an open-source software stack for ADS applications. It encompasses functionalities such as localization, perception, prediction, and planning without delving into specific module implementations. Autoware supports features like valet parking, shuttle buses and cargo delivery. Its modular architecture facilitates scalability and the integration of future enhancements, leveraging ROS 2’s real-time processing and robust communication capabilities.

Baidu's Apollo \cite{baiduApolloOpenAutonomous} runs on a middleware called Cyber RT \cite{baiduIntroductionCyberRT}. Similar to ROS 2, Cyber RT is based on a publish-subscribe model but operates without a central core component, with each module running independently. It uses Google Protocol Buffers for efficient data serialization. In contrast to ROS2, Apollo's middleware is tailored for automated driving, integrating with tools for mapping, localization, traffic management, and vehicle control. Both ROS 2 and Cyber RT provide robust software development kits but differ in approach. Studies suggest Apollo's modules are more robust, though this deviates from the broader ROS2 ecosystem \cite{rajuPerformanceOpenAutonomous2019}. Conversely, ROS2's flexibility and extensive community support make it adaptable for various AV applications, especially in research and development.

Other middleware such as ZeroMQ \cite{ZeroMQ} are used in ADS development \cite{novickisFunctionalArchitectureAutonomous2020}. However, in our view,  ROS2 presents advantages by offering specialized real-time communication and reliability essential for AV safety and performance. Unlike ZeroMQ, which requires additional layers to achieve similar real-time capabilities, ROS2 provides built-in deterministic message delivery and robust quality of service settings. Additionally, ROS2’s ecosystem, supported by the community, accelerates prototyping, allowing developers to quickly integrate and test complex AV systems. 

Both Autoware and Apollo provide a comprehensive environment for ADS development. While these ecosystems are powerful for full-scale AV development, their complexity can hinder the rapid testing and validation of smaller proofs of concept, as they are designed for complete vehicle deployment \cite{rajuPerformanceOpenAutonomous2019}. Their extensive feature sets and deployment capabilities may introduce unnecessary overhead for early-stage development, where simplicity and agility are paramount.

\subsection{Service-Oriented Architecture}

As opposed to traditional monolithic architectures, Service-Oriented Architecture (SOA)  is a paradigm that employs discrete software components, or services, to assemble system applications. Each service encapsulates a distinct functionality, enabling interoperability across diverse platforms and programming languages. SOA facilitates the reuse of services across multiple systems or the integration of independent services to execute more complex tasks. This approach has been recognized by industry experts over several years \cite{10011506} \cite{7930217}. Its benefits have been empirically validated not only for Advanced Driver Assistance Systems (ADAS) functionalities \cite{8712376} but also across other domains \cite{8354415}.

In the automotive context, building on the SOA principles discussed above, services are represented as topics within a data-centric publisher-subscriber model using the DDS, the middleware framework used by ROS2. This structure has been effectively implemented in containerized environments, providing a highly controllable and configurable environment \cite{8417118}.


\subsection{Simulation Platform}

Simulation platforms are an important part of the ADS development lifecycle, providing virtual environments for testing and refining vehicle behaviors without the substantial risks and costs associated with real-world trials. These platforms enable developers to model and evaluate vehicle behavior under diverse conditions, from everyday traffic scenarios to rare and extreme events that are challenging to replicate physically. Leveraging simulations allows for rapid iteration, early issue identification, and optimization of AV algorithms in a controlled and repeatable setting.

The evolution of simulators has seen distinct phases. Initially, from the 1990s to the early 2000s, the focus was primarily on simulating vehicle dynamics, leading to the creation of widely adopted tools such as CarSim \cite{MechanicalSimulationCorporation} and IPG CarMaker \cite{CarMakerIPGAutomotive}. These early platforms enabled detailed modeling of vehicle mechanics but were limited in their ability to simulate complex driving scenarios. The 2000-2015 period brought more sophisticated simulators, like rFpro \cite{AutomotiveSimulationDriving}, a critical tool for testing advanced driver assistance systems (ADAS) and autonomous driving technologies. Since 2015, a surge of open-source simulators has transformed AV research, driven by the demand for high-fidelity and versatile simulation platforms. CARLA \cite{pmlr-v78-dosovitskiy17a} has become one of the most widely adopted platforms, valued for its flexibility and compatibility with plugins like SUMO \cite{SUMO2018} for traffic simulation. Yet, CARLA's limitations in V2X communication simulations have led researchers to supplement it with network-focused simulators.

To address these gaps, several specialized simulators have emerged. Veins \cite{sommerVeinsOpenSource2019}, integrated with OMNeT++ \footnote{https://omnetpp.org/}, focuses on vehicular network (VANET) simulations, making it ideal for Connected Autonomous Vehicle (CAV) applications where communication fidelity is essential. Similarly, Eclipse MOSAIC \cite{schrabModelingITSManagement2023} offers a multi-domain approach, integrating SUMO for traffic flow and ns-3\footnote{https://www.nsnam.org/} for network simulation.


Given its open-source nature, community outreach and developing environment, we selected CARLA as an initial target simulator. However, APIKS service-oriented architecture is agnostic to the simulating platforms through a bridge system. This means, ROS2 messages exchanged within APIKS are not from the nodes exposed by the simulators, but rather internal.

\subsection{ISO/TR 4804}

ISO/TR 4804:2020 \cite{isoRoadVehiclesSafety2020} provides detailed guidelines for ensuring the safety and cybersecurity of automated driving systems. The technical report emphasizes a systematic approach to the design, verification, and validation processes, aiming to guarantee reliable ADS operation across diverse scenarios. By employing a risk-based methodology, ISO/TR 4804 advocates for the early identification and mitigation of potential hazards throughout the development lifecycle.

The report integrates safety and cybersecurity considerations into the fundamental architecture of ADS, aligning with standards such as ISO 26262 \cite{isoRoadVehiclesFunctional} for functional safety and ISO/SAE 21434 \cite{isoRoadVehiclesCybersecurity} for cybersecurity engineering. This integration facilitates a cohesive framework that addresses the multifaceted challenges inherent in automated driving technologies. Key aspects include modular design principles, rigorous testing protocols, and the incorporation of cybersecurity measures from the outset. The principles outlined in the report support the development of platforms with adaptable architectures that can accommodate continuous updates and enhancements without compromising safety or security.

We chose to follow the architectural depiction outlined in the standard because of its emphasis on modularity, which significantly enhances safety. This modularity favors a design where individual components can be developed, tested, and validated independently, reducing the risk of systematic failures. It also allows for the seamless integration of software components, which is particularly relevant for SDVs.

\section{APIKS Automotive Platform} \label{sec:framework}

\begin{figure*}[!t] \centering \includegraphics[width=0.68\textwidth]{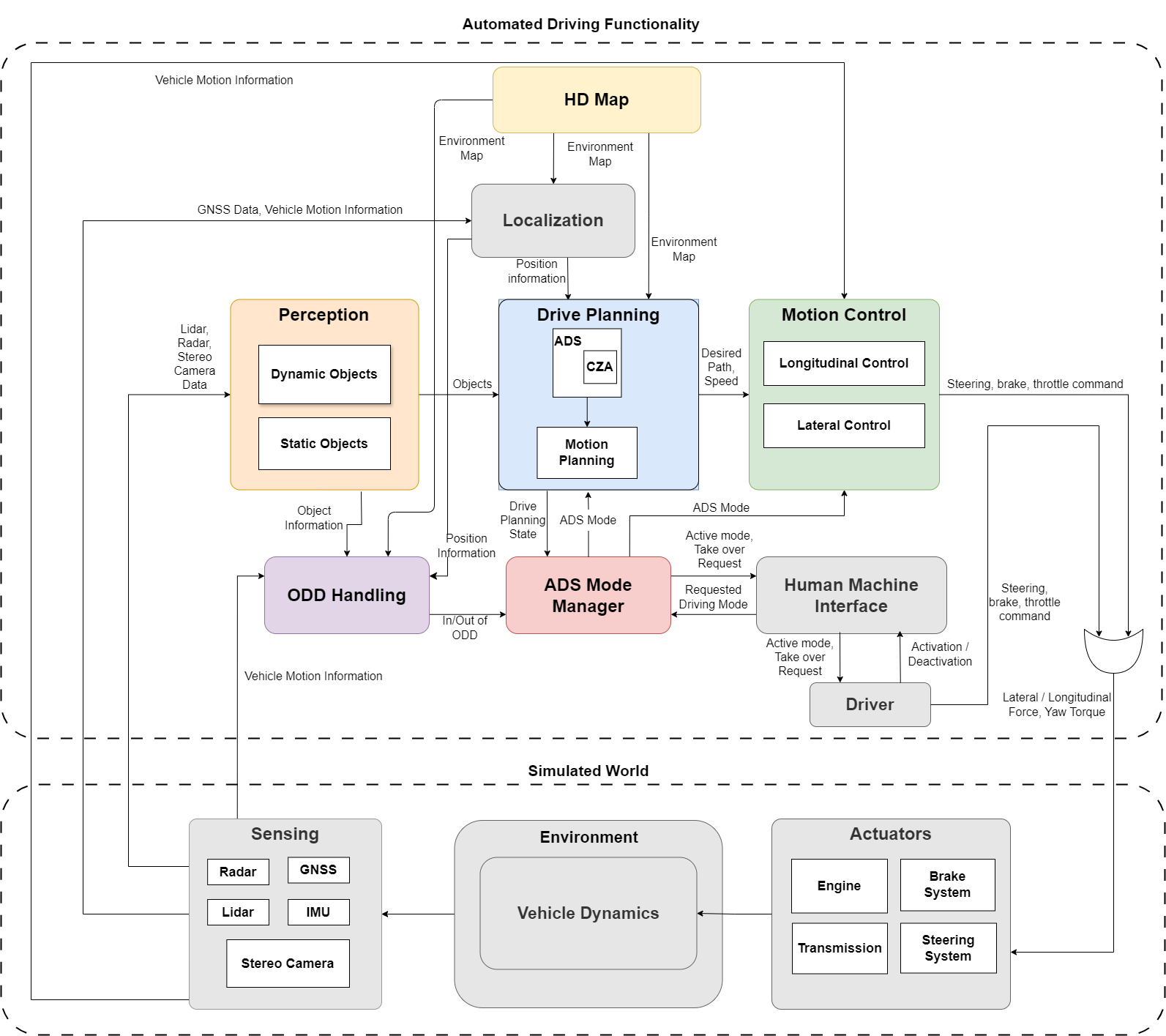} 
\vspace{-0.3cm} 
\caption{Logical architecture of the APIKS platform} 
\vspace{-0.3cm} 
\label{fig:apiks} 
\end{figure*}

The APIKS automotive platform, depicted in Figure \ref{fig:apiks}, is structured into a SOA composed of multiple interconnected layers, each responsible for distinct functions essential to AV operations. This layered approach facilitates modular development and seamless integration, ensuring that each component can be independently developed and maintained while contributing to the overall system's efficiency and safety. At the core, the \emph{sensing Layer} gathers raw data from various sensors, which is then processed by the \emph{perception layer} to build an accurate representation of the vehicle’s environment. This information is further refined by the \emph{operational design domain (ODD) handling layer}, which adjusts operational parameters based on real-time conditions.

Building on this foundation, the \emph{drive planning layer} formulates safe and effective trajectories, which are executed by the \emph{motion control layer} and the \emph{actuation layer} to manage the vehicle's movements. Overseeing these processes, the \emph{ADS mode manager} ensures that the vehicle operates within appropriate modes, adapting to changing scenarios and driver inputs. The \emph{human-machine interface (HMI) layer} provides essential communication between the vehicle and the driver, while the \emph{monitoring and data collection} components enable continuous system evaluation and improvement. In the following subsections, we will delve into each of these layers in detail.

\subsection{High Definition Map Layer} The High Definition (HD) Map provides precise spatial information critical for vehicle localization and trajectory planning. APIKS employs Lanelet2 \cite{poggenhansLanelet2HighdefinitionMap2018}, an open-source mapping framework, to develop and maintain these HD maps. Lanelet2 effectively represents road networks, including lane boundaries, traffic signs, signals, and other infrastructure elements, ensuring detailed and accurate map data.

\subsection{Sensing Layer} The Sensing Layer is responsible for acquiring raw data from an array of sensors, including LiDAR, radar, and cameras. This layer performs initial data processing to extract pertinent features such as object detections and environmental cues, providing foundational inputs for subsequent perception and planning modules.

\subsection{Perception Layer} Building on the processed sensor data, the Perception Layer receives inputs such as images from cameras and point clouds from LiDAR sensors. It processes this data to produce outputs like 2D and 3D object lists, which detail the positions and classifications of detected objects. These outputs collectively create a map of the surrounding environment, providing the vehicle with an accurate and real-time understanding of its operational surroundings.
\begin{figure*}[ht]
    \centering
    \includegraphics[width=0.7\textwidth]{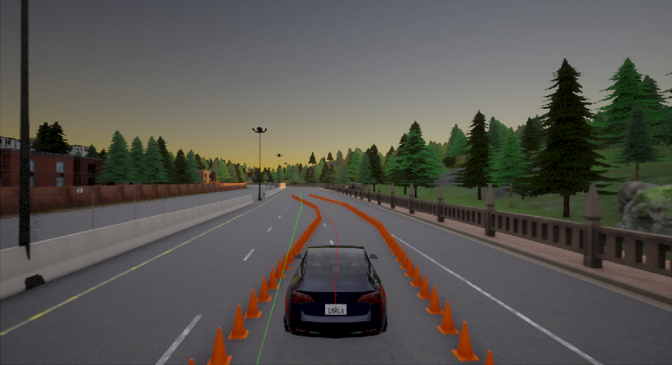}
    \vspace{-0.3cm}
    \caption{Example scenario of Construction Zone Assist use case.}
    \vspace{-0.3cm}
    \label{fig:cza}
\end{figure*}
\subsection{Operational Design Domain Handling Layer} The ODD Handling Layer dynamically manages the vehicle's operational parameters by interpreting real-time data from various sources. ODD messages are modeled in accordance with ISO 34503 \cite{isoRoadVehiclesTest2023}, ensuring standardized communication and interoperability across different systems. By evaluating current driving conditions—including road boundaries, traffic scenarios, and the presence of dynamic obstacles—this layer enables the vehicle to adjust its behavior within its defined operational domain. Additionally, the ODD handler facilitates the Automated Driving System (AS) mode manager in determining the availability of specific functionalities based on the current ODD. For instance, adverse weather conditions such as heavy rainfall may reduce visibility, prompting the system to disable autopilot features or initiate a handover to the driver.

\subsection{Drive Planning Layer} The Drive Planning Layer is tasked with generating safe and feasible trajectories based on the vehicle's environmental understanding. It integrates data from the Perception and ODD Handling Layers to navigate complex environments while avoiding obstacles and adhering to traffic regulations. This layer can employ various methodologies to achieve robust and adaptable planning. For example, ground-truth-based planning involves following a pre-calculated global path with comprehensive knowledge of surrounding objects, ensuring precise adherence to the planned route under known conditions. Additionally, other drive planning methods tested within APIKS include Model Predictive Control (MPC) \cite{richaletModelPredictiveHeuristic1978} and integration with TUM's Frenetix framework \cite{trauthFRENETIXHighPerformanceModular2024}. MPC optimizes trajectories in real time by continuously predicting and adjusting the vehicle's path, allowing for responsive maneuvering in unpredictable environments. The Frenetix framework utilizes a Frenet \cite{werlingOptimalTrajectoryGeneration2010} path planner to generate smooth and efficient paths relative to a reference trajectory, simplifying complex path planning by transforming it into longitudinal and lateral components. The use of these diverse planning approaches underscores the architecture's modularity: as long as the interfaces are properly defined and adhered to, different planning algorithms can be developed and integrated seamlessly.

\subsection{Motion Control Layer} Translating planned trajectories into executable commands, the Motion Control Layer employs a decoupled architecture that independently manages longitudinal and lateral vehicle dynamics. Longitudinal control is managed by a Proportional-Integral-Derivative (PID) controller, which precisely regulates acceleration and braking to maintain desired speed profiles and respond to dynamic conditions. For lateral control, the Stanley controller is utilized. The Stanley controller is designed to minimize both the cross-track error, which is the lateral deviation from the planned path, and the heading error, which is the angular difference between the vehicle's orientation and the desired trajectory. By addressing these two error components, the Stanley controller ensures accurate steering adjustments that align the vehicle with the intended path. This separation of control functions allows each controller to specialize and optimize its specific task. 

\subsection{Actuation Layer} The Actuation Layer interfaces directly with the vehicle's hardware components, executing control commands issued by the Motion Control Layer. It manages systems such as the engine, brakes, and steering mechanisms, ensuring that the vehicle's physical actions correspond accurately to the planned trajectories and control inputs.

\subsection{ADS Mode Manager} The Automated Driving System Mode Manager is responsible for overseeing the operational modes of the vehicle by managing the activation and deactivation of autonomous driving functionalities. It ensures that the system operates in modes that are appropriate to the current context and aligned with the driver’s intentions, thereby facilitating seamless transitions between manual and autonomous control as required. The Mode Manager interacts closely with other system components, such as the ODD handler, to evaluate environmental conditions and system status. Additionally, the Mode Manager coordinates with the Drive Planner, requesting modifications to the planning algorithms based on prevailing conditions. For instance, navigating a construction zone may necessitate a specialized drive planner that accounts for temporary road structures and altered traffic patterns, which differs from the drive planner optimized for highway environments. Furthermore, the Mode Manager continuously monitors the performance and status of various system modules, enabling it to detect anomalies or critical situations promptly. In scenarios requiring immediate action, such as initiating emergency braking, the Mode Manager can override standard procedures and bypass components like the Drive Planner to execute necessary emergency maneuvers.

\subsection{Human-Machine Interface Layer} The Human-Machine Interface (HMI) Layer serves as the communication bridge between the autonomous system and the driver. It provides essential feedback, including system status updates, warnings, and take-over requests, enabling the driver to remain informed and intervene when required.

\subsection{Monitoring and Data Collection} By leveraging ROS2’s publish–subscribe architecture, monitoring and data collection can be seamlessly integrated into the system for real-time observation and data acquisition. Although APIKS does not currently include built-in data collection and monitoring capabilities, its design makes it easy to implement these features. This enables post-operation analysis and continuous system improvement by gathering data on algorithm performance and system effectiveness across various operating conditions, thereby supporting informed enhancements to the platform.
\section{Construction Zone Assist}
\label{sec:cza}

To demonstrate APIKS, we deployed a widely relevant application for AV: the \emph{Construction Zone Assist} (CZA) system. Figure This automated functionality is conceived to detect highway construction zones with a high degree of safety and reliability. Construction zones present unique challenges due to their unpredictable and dynamic nature, which may include irregular road geometries, temporary signage, and a variety of static and dynamic obstacles such as construction machinery and workers. By leveraging depth camera technology and YOLOv8 \cite{ultralytics2023yolov8} for object detection, the CZA system enables automated vehicles to accurately perceive and navigate these complex environments.

The system is based on the premise that an existing automated driving feature, such as an \emph{autopilot}, is already integrated into the vehicle. When the vehicle operates in this automated mode and a construction zone is detected, the system assesses whether the current ODD permits a mode transition. If the conditions are favourable, the system switches to an alternative mode to activate the appropriate functionality for navigating the construction zone. The primary modifications occur within the vehicle's drive planning algorithms, tailored to navigate the static objects limiting the route, constraint target velocities and safety distances. This is supported by using Frenetix \cite{trauthFRENETIXHighPerformanceModular2024} to plan the trajectories according to \cite{werlingOptimalTrajectoryGeneration2010}. 

Figure \ref{fig:cza} illustrates a scenario demonstrating the operation of the CZA functionality. The green line represents the initially computed target path without knowledge of the temporary deviation. The red line depicts the real-time computed path that allows navigation through the construction zone. This example was deployed and tested using the APIKS platform connected to the CARLA simulator via its ROS-bridge. Additionally, we utilized the scenario runner feature, which allows us to abstract the testing of the functionality using a standardized format—namely, OpenSCENARIO from ASAM. This scenario demonstrates new functionalities are able to be deployed using previously developed modules within the APIKS platform to adapt the vehicle's path in real-time. It shows how the system effectively navigates through temporary deviations like construction zones, validating the functionality of the CZA feature.

\section{Conclusion}
\label{sec:conclusion}

This paper introduced APIKS, a modular platform based on a service-oriented architecture for early-stage automated driving systems development and testing. Utilizing the ROS2 middleware, APIKS integrates key functions in sensing, perception, planning, and control, enabling rapid prototyping and allowing developers to iterate quickly without the complexity of full-scale systems. Its modular design and software-defined features provide flexibility, making it advantageous over existing simulation and SDK platforms. APIKS adheres to industry standards like ISO/TR 4804 and was demonstrated in a construction zone assist scenario, highlighting its capability to manage complex, dynamic environments. Future work will expand its operational domain, incorporate other automated functionalities, and add driver monitoring capabilities by integrating the driver into the system, further supporting the development of safe and reliable automated driving systems.

\bibliographystyle{IEEEtran}
\bibliography{bibliography}

\begin{thebibliography}{10}
\providecommand{\url}[1]{#1}
\csname url@samestyle\endcsname
\providecommand{\newblock}{\relax}
\providecommand{\bibinfo}[2]{#2}
\providecommand{\BIBentrySTDinterwordspacing}{\spaceskip=0pt\relax}
\providecommand{\BIBentryALTinterwordstretchfactor}{4}
\providecommand{\BIBentryALTinterwordspacing}{\spaceskip=\fontdimen2\font plus
\BIBentryALTinterwordstretchfactor\fontdimen3\font minus
  \fontdimen4\font\relax}
\providecommand{\BIBforeignlanguage}[2]{{%
\expandafter\ifx\csname l@#1\endcsname\relax
\typeout{** WARNING: IEEEtran.bst: No hyphenation pattern has been}%
\typeout{** loaded for the language `#1'. Using the pattern for}%
\typeout{** the default language instead.}%
\else
\language=\csname l@#1\endcsname
\fi
#2}}
\providecommand{\BIBdecl}{\relax}
\BIBdecl

\bibitem{wangQuantificationSafetyImprovements2024}
S.~Wang, Z.~Li, Y.~Wang, W.~Zhao, and H.~Wei, ``Quantification of safety
  improvements and human-machine tradeoffs in the transition to automated
  driving,'' vol. 199, p. 107523.

\bibitem{fengDenseReinforcementLearning2023}
\BIBentryALTinterwordspacing
S.~Feng, H.~Sun, X.~Yan, H.~Zhu, Z.~Zou, S.~Shen, and H.~X. Liu, ``Dense
  reinforcement learning for safety validation of autonomous vehicles,''
  \emph{Nature}, vol. 615, no. 7953, pp. 620--627, Mar. 2023. [Online].
  Available: \url{https://doi.org/10.1038/s41586-023-05732-2}
\BIBentrySTDinterwordspacing

\bibitem{baiduApolloOpenAutonomous}
\BIBentryALTinterwordspacing
Baidu, ``Apollo: {An} open autonomous driving platform.'' [Online]. Available:
  \url{https://github.com/ApolloAuto/apollo}
\BIBentrySTDinterwordspacing

\bibitem{autowarefoundationAutoware}
\BIBentryALTinterwordspacing
A.~Foundation, ``\BIBforeignlanguage{en-US}{Autoware}.'' [Online]. Available:
  \url{https://autoware.org/}
\BIBentrySTDinterwordspacing

\bibitem{macenskiRobotOperatingSystem2022}
\BIBentryALTinterwordspacing
S.~Macenski, T.~Foote, B.~Gerkey, C.~Lalancette, and W.~Woodall, ``Robot
  {Operating} {System} 2: {Design}, {Architecture}, and {Uses} {In} {The}
  {Wild},'' Nov. 2022, arXiv:2211.07752. [Online]. Available:
  \url{http://arxiv.org/abs/2211.07752}
\BIBentrySTDinterwordspacing

\bibitem{isoRoadVehiclesSafety2020}
\BIBentryALTinterwordspacing
ISO, ``Road vehicles — {Safety} and cybersecurity for automated driving
  systems — {Design}, verification and validation,'' 2020. [Online].
  Available: \url{https://www.iso.org/standard/80363.html}
\BIBentrySTDinterwordspacing

\bibitem{pardo-castelloteOMGDataDistributionService2003}
\BIBentryALTinterwordspacing
G.~Pardo-Castellote, ``{OMG} {Data}-{Distribution} {Service}: architectural
  overview,'' in \emph{23rd {International} {Conference} on {Distributed}
  {Computing} {Systems} {Workshops}, 2003. {Proceedings}.}, May 2003, pp.
  200--206. [Online]. Available:
  \url{https://ieeexplore.ieee.org/document/1203555}
\BIBentrySTDinterwordspacing

\bibitem{quigleyROSOpensourceRobot}
M.~Quigley, B.~Gerkey, K.~Conley, J.~Faust, T.~Foote, J.~Leibs, E.~Berger,
  R.~Wheeler, and A.~Ng, ``\BIBforeignlanguage{en}{{ROS}: an open-source
  {Robot} {Operating} {System}}.''

\bibitem{baiduIntroductionCyberRT}
\BIBentryALTinterwordspacing
Baidu, ``\BIBforeignlanguage{en}{Introduction into {Cyber} {RT}}.'' [Online].
  Available:
  \url{https://github.com/fortiss/apollo/blob/r5.5.0/cyber/README.md}
\BIBentrySTDinterwordspacing

\bibitem{rajuPerformanceOpenAutonomous2019}
\BIBentryALTinterwordspacing
V.~M. Raju, V.~Gupta, and S.~Lomate, ``Performance of {Open} {Autonomous}
  {Vehicle} {Platforms}: {Autoware} and {Apollo},'' in \emph{2019 {IEEE} 5th
  {International} {Conference} for {Convergence} in {Technology} ({I2CT})},
  Mar. 2019, pp. 1--5. [Online]. Available:
  \url{https://ieeexplore.ieee.org/document/9033734/?arnumber=9033734}
\BIBentrySTDinterwordspacing

\bibitem{ZeroMQ}
\BIBentryALTinterwordspacing
``{ZeroMQ}.'' [Online]. Available: \url{https://zeromq.org/}
\BIBentrySTDinterwordspacing

\bibitem{novickisFunctionalArchitectureAutonomous2020}
\BIBentryALTinterwordspacing
R.~Novickis, A.~Levinskis, R.~Kadikis, V.~Fescenko, and K.~Ozols, ``Functional
  {Architecture} for {Autonomous} {Driving} and its {Implementation},'' in
  \emph{2020 17th {Biennial} {Baltic} {Electronics} {Conference} ({BEC})}, Oct.
  2020, pp. 1--6, iSSN: 2382-820X. [Online]. Available:
  \url{https://ieeexplore.ieee.org/document/9276943/?arnumber=9276943}
\BIBentrySTDinterwordspacing

\bibitem{10011506}
N.~Kukulicic, D.~Samardzic, A.~Bucaioni, and S.~Mubeen, ``Automotive
  service-oriented architectures: a systematic mapping study,'' in \emph{2022
  48th euromicro conference on software engineering and advanced applications
  ({SEAA})}, 2022, pp. 459--466.

\bibitem{7930217}
S.~Kugele, P.~Obergfell, M.~Broy, O.~Creighton, M.~Traub, and W.~Hopfensitz,
  ``On service-orientation for automotive software,'' in \emph{2017 {IEEE}
  international conference on software architecture ({ICSA})}, 2017, pp.
  193--202.

\bibitem{8712376}
J.~Lotz, A.~Vogelsang, O.~Benderius, and C.~Berger, ``Microservice
  architectures for advanced driver assistance systems: a case-study,'' in
  \emph{2019 {IEEE} international conference on software architecture companion
  ({ICSA}-c)}, 2019, pp. 45--52.

\bibitem{8354415}
A.~Bucchiarone, N.~Dragoni, S.~Dustdar, S.~T. Larsen, and M.~Mazzara, ``From
  monolithic to microservices: {An} experience report from the banking
  domain,'' \emph{IEEE Software}, vol.~35, no.~3, pp. 50--55, 2018.

\bibitem{8417118}
S.~Kugele, D.~Hettler, and J.~Peter, ``Data-centric communication and
  containerization for future automotive software architectures,'' in
  \emph{2018 {IEEE} international conference on software architecture
  ({ICSA})}, 2018, pp. 65--6509.

\bibitem{MechanicalSimulationCorporation}
\BIBentryALTinterwordspacing
``Mechanical {Simulation} {Corporation}.'' [Online]. Available:
  \url{https://www.carsim.com/}
\BIBentrySTDinterwordspacing

\bibitem{CarMakerIPGAutomotive}
\BIBentryALTinterwordspacing
``{CarMaker} {\textbar} {IPG} {Automotive}.'' [Online]. Available:
  \url{https://www.ipg-automotive.com/en/products-solutions/software/carmaker/}
\BIBentrySTDinterwordspacing

\bibitem{AutomotiveSimulationDriving}
\BIBentryALTinterwordspacing
``\BIBforeignlanguage{en-GB}{Automotive {Simulation} • {Driving} {Simulation}
  • {Autonomous} {Driving} • {rFpro}}.'' [Online]. Available:
  \url{https://rfpro.com/}
\BIBentrySTDinterwordspacing

\bibitem{pmlr-v78-dosovitskiy17a}
\BIBentryALTinterwordspacing
A.~Dosovitskiy, G.~Ros, F.~Codevilla, A.~Lopez, and V.~Koltun, ``{CARLA}: {An}
  open urban driving simulator,'' in \emph{Proceedings of the 1st annual
  conference on robot learning}, ser. Proceedings of machine learning research,
  S.~Levine, V.~Vanhoucke, and K.~Goldberg, Eds., vol.~78.\hskip 1em plus 0.5em
  minus 0.4em\relax PMLR, Nov. 2017, pp. 1--16. [Online]. Available:
  \url{https://proceedings.mlr.press/v78/dosovitskiy17a.html}
\BIBentrySTDinterwordspacing

\bibitem{SUMO2018}
P.~A. Lopez, M.~Behrisch, L.~Bieker-Walz, J.~Erdmann, Y.-P. Fl{\"o}tter{\"o}d,
  R.~Hilbrich, L.~L{\"u}cken, J.~Rummel, P.~Wagner, and E.~Wie{\ss}ner,
  ``Microscopic traffic simulation using sumo,'' in \emph{2018 21st
  International Conference on Intelligent Transportation Systems (ITSC)}.\hskip
  1em plus 0.5em minus 0.4em\relax IEEE, 2018, pp. 2575--2582.

\bibitem{sommerVeinsOpenSource2019}
C.~Sommer, D.~Eckhoff, A.~Brummer, D.~S. Buse, F.~Hagenauer, S.~Joerer, and
  M.~Segata, ``Veins: {The} {Open} {Source} {Vehicular} {Network} {Simulation}
  {Framework},'' in \emph{Recent {Advances} in {Network} {Simulation}}.\hskip
  1em plus 0.5em minus 0.4em\relax Springer International Publishing, 2019, pp.
  215--252.

\bibitem{schrabModelingITSManagement2023}
K.~Schrab, M.~Neubauer, R.~Protzmann, I.~Radusch, S.~Manganiaris, P.~Lytrivis,
  and A.~J. Amditis, ``Modeling an {ITS} {Management} {Solution} for {Mixed}
  {Highway} {Traffic} {With} {Eclipse} {MOSAIC},'' \emph{Trans. Intell.
  Transport. Sys.}, vol.~24, no.~6, pp. 6575--6585, Jun. 2023.

\bibitem{isoRoadVehiclesFunctional}
\BIBentryALTinterwordspacing
ISO, ``Road vehicles — {Functional} safety.'' [Online]. Available:
  \url{https://www.iso.org/standard/68383.html}
\BIBentrySTDinterwordspacing

\bibitem{isoRoadVehiclesCybersecurity}
ISO and SAE, ``Road vehicles — {Cybersecurity} engineering.''

\bibitem{poggenhansLanelet2HighdefinitionMap2018}
\BIBentryALTinterwordspacing
F.~Poggenhans, J.-H. Pauls, J.~Janosovits, S.~Orf, M.~Naumann, F.~Kuhnt, and
  M.~Mayr, ``Lanelet2: {A} high-definition map framework for the future of
  automated driving,'' in \emph{2018 21st {International} {Conference} on
  {Intelligent} {Transportation} {Systems} ({ITSC})}, Nov. 2018, pp.
  1672--1679, iSSN: 2153-0017. [Online]. Available:
  \url{https://ieeexplore.ieee.org/document/8569929}
\BIBentrySTDinterwordspacing

\bibitem{isoRoadVehiclesTest2023}
\BIBentryALTinterwordspacing
{ISO}, ``\BIBforeignlanguage{en}{Road {Vehicles} — {Test} scenarios for
  automated driving systems {Specification} for operational design domain},''
  2023. [Online]. Available: \url{https://www.iso.org/standard/78952.html}
\BIBentrySTDinterwordspacing

\bibitem{richaletModelPredictiveHeuristic1978}
J.~Richalet, A.~Rault, J.~L. Testud, and J.~Papon, ``Model predictive heuristic
  control: {{Applications}} to industrial processes,'' vol.~14, no.~5, pp.
  413--428.

\bibitem{trauthFRENETIXHighPerformanceModular2024}
R.~Trauth, K.~Moller, G.~Würsching, and J.~Betz, ``{{FRENETIX}}: {{A
  High-Performance}} and {{Modular Motion Planning Framework}} for {{Autonomous
  Driving}},'' vol.~12, pp. 127\,426--127\,439.

\bibitem{werlingOptimalTrajectoryGeneration2010}
M.~Werling, J.~Ziegler, S.~Kammel, and S.~Thrun, ``Optimal trajectory
  generation for dynamic street scenarios in a {{Frenét Frame}},'' in
  \emph{2010 {{IEEE International Conference}} on {{Robotics}} and
  {{Automation}}}.\hskip 1em plus 0.5em minus 0.4em\relax IEEE, pp. 987--993.

\bibitem{ultralytics2023yolov8}
Ultralytics, ``Yolov8,'' \url{https://github.com/ultralytics/ultralytics},
  2023.

\end{thebibliography}

\end{document}